\documentclass[epj]{svjour}
\usepackage{graphics}
\def\Journal#1#2#3#4{{#1} {\bf #2}, (#4) #3}
\def\EPJ{Eur.~Phys.~J}
\def\JHEP{Journal High Energy Phys.}

\def\NPB{Nucl. Phys. B}

\def\PLB{Phys. Lett. B}
\def\PRL{Phys. Rev. Lett.}
\def\PRD{Phys. Rev. D}
\def\ZPC{Z. Phys. C}
\begin{document}
\title{Quark-Hadron Duality in Structure Functions}
\author{Alessandra Fantoni\inst{1}, Nicola Bianchi\inst{1} 
\and Simonetta Liuti\inst{2}
}                     
\offprints{A. Fantoni} 
\institute{Laboratori Nazionali di Frascati dell'INFN, 
Via E. Fermi 40, 00044 Frascati (RM), Italy
\and 
University of Virginia, Charlottesville, Virginia 22901, USA}
\date{Received: date / Revised version: date}
%
\abstract{
The quark-hadron duality is studied in a systematic way for polarized and
unpolarized structure functions, by taking into account all the available data 
in the resonance region.
In both cases, a precise perturbative QCD based analysis to the integrals
of the structure functions in the resonance region has been done: non perturbative
contributions have been disentangled and the higher twist contributions have
been evaluated. A different behavior for the unpolarized and polarized structure
functions at low $Q^2$ has been found.
\PACS{
      {PACS-key}{quark-hadron duality}   \and
      {PACS-key}{structure functions}
     } 
} 
\maketitle
\section{Introduction}
\label{intro}
Understanding the structure and interaction of hadron in terms of the 
quark and gluon degrees of freedom of QCD is one of the unsolved 
problems of the Standard Model of nuclear and particle physics.
At present it's not possible to describe the physics of hadrons
directly from QCD, however it is known that it should just be a matter
of convenience the choice of describing a process in terms of 
quark-gluon or hadronic degrees of freedom.
This concept is called quark-hadron duality. 
At high energies, where the interactions between quarks and gluons
become weak and quark can be considered asymptotically free, an 
efficient description of phenomena is possible in terms of quarks.
At low energies, where the effects of confinement become large, it
is more efficient to work in terms of collective degrees of freedom, 
the physical mesons and baryons.
In these terms, it's clear that the duality between the quark and
hadron descriptions reflects the relationship between confinement
and asymptotic freedom, and is intimately related to the nature of
the transition from non-perturbative (low energy) to perturbative QCD
(high energy). 

The concept of duality was introduced for the first time by Bloom and 
Gilman \cite{BG} in deep inelastic scattering (DIS).
They noticed an equivalence between the smooth $x$ dependence of the inclusive 
structure function at large $Q^2$ and the average over $W^2$ of the 
nucleon resonances.
Furthermore, this equivalence appeared to hold for each resonance, over 
restricted regions in $W$.
Based on this observations, one can refer to {\em global} duality if the 
average, defined {\it e.g.} as the integral of the structure functions, 
is taken over the whole resonance region $ 1 \leq W^2 \leq 4$ GeV$^2$. 
If, however, the averaging is performed over smaller $W^2$ ranges, extending 
{\it e.g.} over single resonances, one can refer to {\em local} duality. 

Although the duality between quark and hadron descriptions is, in principle,
formally exact, how this reveals itself specifically in different physical
processes and under different kinematical conditions is the key to
understand the consequences of QCD for hadronic structure.
The phenomenon of duality is quite general in nature and can be studied in
a variety of processes, such as DIS, $e^+e^-$ annihilation into hadrons, 
and hadron-hadron collisions, or semi-leptonic decays of heavy quarks.   
With the advent of both more detailed studies of soft scales and 
confinement \cite{reticolo}, and higher precision measurements covering
a wide range of reactions, it is now becoming 
possible to investigate the role of duality in QCD as a 
subject per se. 

\section{Kinematical variables}
\begin{figure}
\resizebox{0.45\textwidth}{!}{%
  \includegraphics{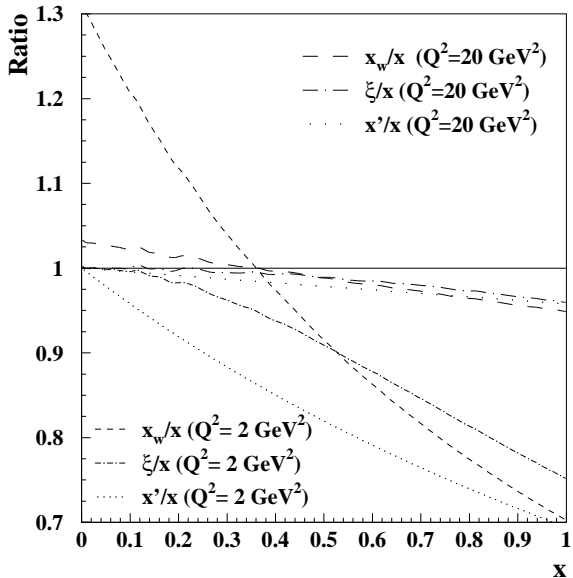}
}
\caption{Ratio between the three variables $x'$, $\xi$ and $x_W$ defined 
         in the text and the Bjorken variable $x$ as a function of $x$.}
\label{variables}      
\end{figure}

Besides the scaling variable $x$, other variables have been used in the 
literature to study duality and a number of parametrizations based on these 
variables have been proposed that reproduce in an effective 
way some of the corrections to the perturbative QCD calculations.
The most extensively used variables are: 
$x'=1/\omega'$, where $\omega'= 1/x + M^2/Q^2$, originally introduced by 
Bloom and Gilman in order to obtain a better agreement between DIS and the 
resonance region; 
$\xi=2x/(1+(1+4x^2M^2/Q^2)^{1/2})$ \cite{NAC}, 
originally introduced to take into account the target mass 
effects; 
$x_w=Q^2+B/(Q^2+W^2-M^2+A)$, $A$ and $B$ being fitted parameters,  
used in Refs.~\cite{SzcUle,BOYA}. 
These additional variables include 
a $Q^2$ dependence that phenomenologically absorbs some of
the scaling violations that are important  
at low $Q^2$.
In Fig.\ref{variables}  
their behavior vs. $x$ is compared  for different values of $Q^2$.
From the figure one can see that by calculating $F_2$ in $\xi$ and $x'$, 
one effectively ``rescales'' the structure function to lower values of $x$,
in a $Q^2$ dependent way, namely the rescaling is larger at lower $Q^2$.  
In the analysis reported in the following, the $x$ variable has been
used and all the corrections have been applied one by one.

\section{Analysis of data}
A quantitative analysis of the $Q^2$ dependence of quark-hadron duality 
in both polarized and unpolarized $ep$ scattering is presented. 
All current data in the resonance region, $1 \leq W^2 \leq 4$ GeV$^2$,
have been taken into account.
For the unpolarized case it has been used the data obtained at Jefferson 
Lab in the range $0.3 \leq Q^2 \leq 5$ GeV$^2$ \cite{CEBAF}, and the data 
from SLAC (\cite{whit} and references therein) for $Q^2 \geq 4$ GeV$^2$.
For the polarized case there are only few experimental data in the 
resonance region. 
One set is part of the E143 data \cite{E143}, and it corresponds to 
$Q^2 =0.5$ and $1.2$ GeV$^2$. 
Another set is the one from HERMES \cite{HERMESPRL,Ale} in the range
$1.2 \leq Q^2 \leq 12$ GeV$^2$. 

In the polarized case the $Q^2$ dependence originates from 
the structure function $F_1$ and from the ratio $R$, for which
the SLAC global analysis~\cite{r1990} parametrization has been
used. 
For the asymmetry $A_1$, it was used a power law fit to the world DIS 
data at $x>$0.3, $A_1=x^{0.7}$, as already shown in Ref.~\cite{HERMESPRL}.
This parametrization of $A_1$ is constrained to 1 at $x$=1 and it does not
depend on $Q^2$, as indicated by experimental data in this range~\cite{E155}. 

The full procedure of the analysis is described in \cite{BFL}.
The quark-hadron duality in DIS is studied by considering the ratio of the
integrals of the structure functions integrated in a defined $x$-range,
corresponding to the $W$ range of the resonance region. The structure function
in the numerator is evaluated using the experimental data in the resonance
region, while the one at the denominator is calculated from parametrizations
that reproduce the DIS behavior of the data at large $Q^2$. 
The ratios have been calculated in unpolarized and polarized cases.
It has been found~\cite{BFL} that quark-hadron duality has not been fulfilled 
by using solely the parton distribution functions up to NLO in both the 
unpolarized and polarized structure functions $F_2$ and $g_1$.
However it was possible to see a different behavior between
$R_{\mathrm{unpol}}$ and $R_{\mathrm{pol}}$.
In the unpolarized case the ratio is increasing with $Q^2$, but for
the polarized case the situation is different: while at low $Q^2$ 
the ratio is significantly below unity and shows a strong increase 
with $Q^2$, at higher $Q^2$ the ratio derived from HERMES is above unity and
it appears to be weakly dependent of $Q^2$ within error bars.
The situation is different with the use of the phenomenological fits to DIS 
data~\cite{ALLM,NMC,BOYA}. 
Since these phenomenological parametrizations are obtained by fitting 
deep-inelastic data even in the low $Q^2$ region, they can implicitly include 
non-perturbative effects and this may explain the ``observation of duality''.
It becomes really important to understand the contribution of these
non-perturbative effects.

\section{Size of non-perturbative contributions}
In order to understand the nature of the remaining $Q^2$ dependence that cannot
be described by NLO pQCD evolution, the effect of target mass corrections and 
large $x$ resummation have been studied. 
As mentioned early, the analysis was performed by using $x$ as an integration 
variable, which avoids the ambiguities associated to the usage of other 
{\it ad hoc} kinematical variables.
Standard input parametrizations with initial scale $Q_o^2 = 1 $ GeV$^2$ have
been used. Once both effects have been subtracted from the data, and assuming 
the validity of the twist expansion, one can interpret any remaining 
discrepancy of the ratio from unity in terms of higher twist.

\subsection{Target Mass Corrections (TMC)}
TMC are necessary to take into account the finite mass
of the initial nucleon. They are corrections to the leading twist (LT) part
of the unpolarized structure function $F_2$.
For $Q^2$ larger than $\approx $ 1 GeV$^2$, TMC are taken into
account through the following expansion \cite{GeoPol}: 
\begin{equation}
\label{TMC}
F_{2}^{TMC}(x,Q^2) = \frac{x^2}{\xi^2\gamma^3}F_2^{\mathrm{\infty}}(\xi,Q^2) + 
6\frac{x^3M^2}{Q^2\gamma^4}\int_\xi^1\frac{d \xi'}{{\xi'}^2}
\end{equation}
where $F_2^{\infty}$ is the structure function in the absence of TMC.  
Following the original suggestion of \cite{Mir}, only terms up to order 
$M^2/Q^2$ are kept in the expansion, so as to minimize  
ambiguities in the behavior of $F_2$ at $x \approx 1$.  
Although this procedure disregards parton off-shell effects that might be 
important in the resonance region~\cite{FraGun,ioanaPRD}, it's important to 
emphasize its power expansion character, and setting as a 
limiting condition for its validity, that the inequality $x^2M^2/Q^2 < 1$ 
be verified \cite{SIMO1}, $Q^2 \simeq$ 1.5 GeV$^2$.

\subsection{Large $x$ Resummation (L$x$R)}
LxR effects arise formally from terms containing powers of $\ln (1-z)$, $z$ 
being the longitudinal variable in the evolution equations, that are present 
in the Wilson coefficient functions $C(z)$. 
The latter relate the parton distributions to {\it e.g.} the structure 
function $F_2$, according to:
\begin{equation}
F_2^{NS}(x,Q^2)  = \frac{\alpha_s}{2\pi} \sum_q \int_x^1 dz \, C_{NS}(z) \, 
                   q_{NS}(x/z,Q^2), 
\end{equation}   
where it has been considered only the non-singlet (NS) contribution to $F_2$ 
since only valence quarks distributions are relevant in the present kinematics.
The logarithmic terms in $C_{NS}(z)$ become very large at large $x$, and they 
need to be resummed to all orders in $\alpha_S$. 
This can be accomplished by noticing that the correct kinematical variable that
determines the phase space for the radiation of gluons at large $x$, 
is $\widetilde{W}^2 = Q^2(1-z)/z$, instead of $Q^2$ \cite{BroLep,Ama}. 
As a result, the argument of the strong coupling constant becomes 
$z$-dependent: 
$\alpha_S(Q^2) \rightarrow \alpha_S(Q^2 (1-z)/z)$ (\cite{Rob} and references 
therein). 
In this procedure, however, an ambiguity is introduced, related to the need of 
continuing the value of $\alpha_S$ for low values of its argument, {\it i.e.} 
for $z$ very close to $1$ \cite{PenRos}. 
The size of this ambiguity could be of the same order of the HT corrections. 
Nevertheless, the present evaluation is largely free from this problem because 
of the particular kinematical conditions in the resonance region. 
In this analysis, in fact, the structure functions have been studied 
at {\it fixed}  $W^2$, in between $1 \leq W^2 \leq 4$ GeV$^2$. 
Consequently $Q^2$ increases with $x$. 
This softens the ambiguity in $\alpha_S$, and renders this procedure reliable 
for the extraction of HT terms. 

\subsection{Disentangle of non-perturbative contributions}
\begin{figure}
\resizebox{0.45\textwidth}{!}{%
  \includegraphics{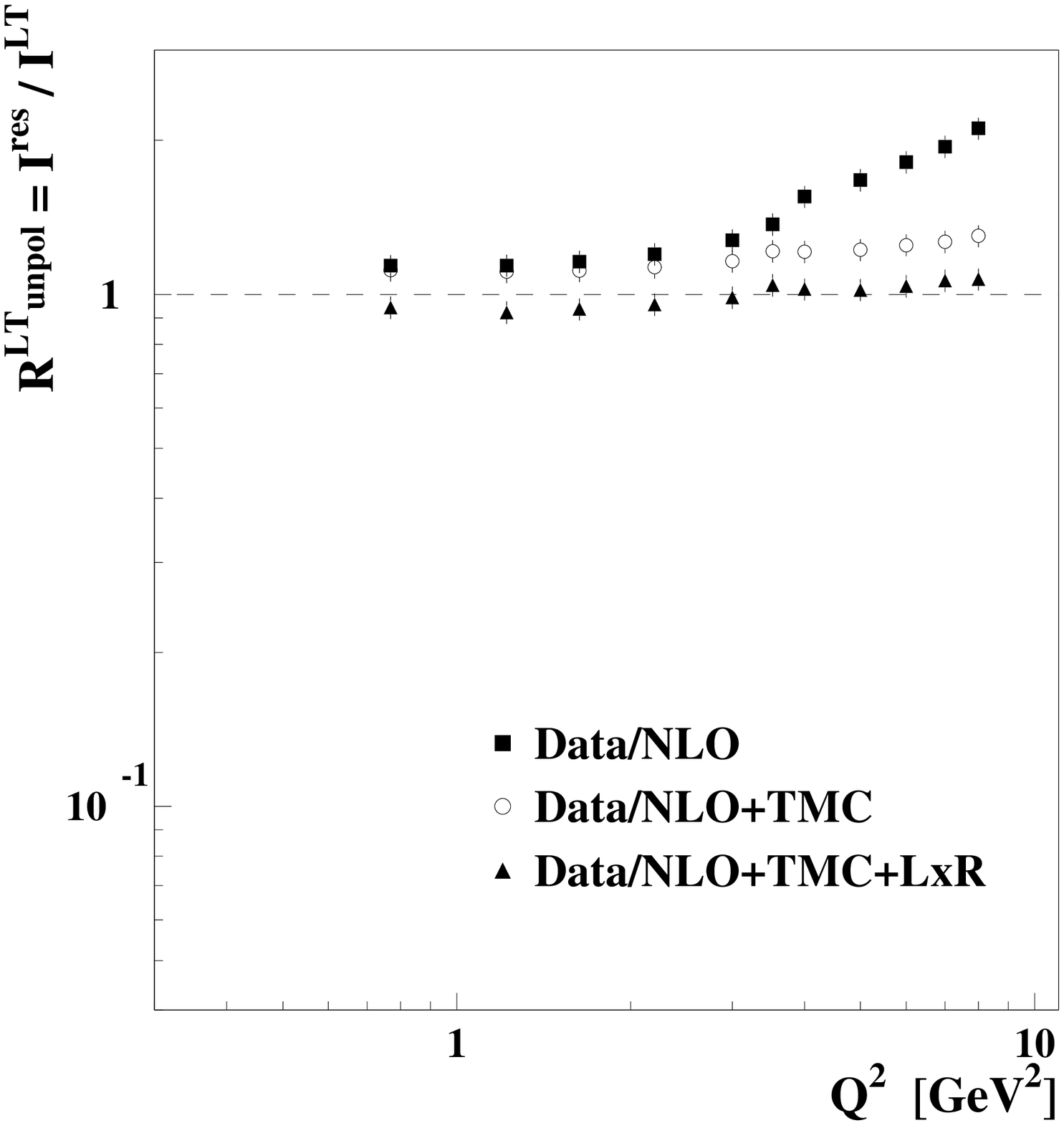}}
\resizebox{0.45\textwidth}{!}{%
  \includegraphics{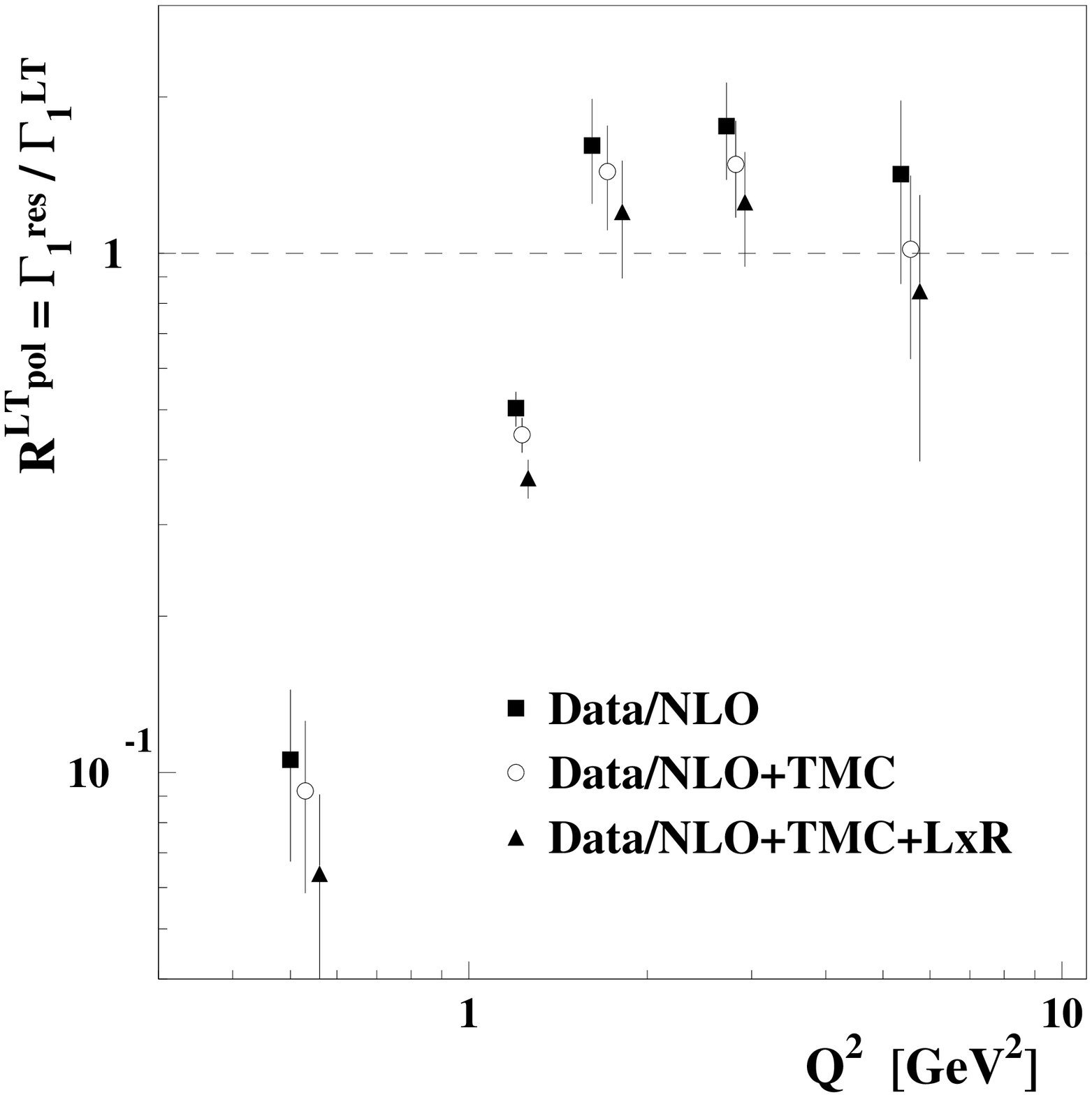}
}
\caption{Ratio between the integrals of the measured structure functions 
         and the calculated ones plotted as a function of $Q^2$.  
         The calculation includes one by one the effects of NLO pQCD
         (squares), TMC (open circles) and LxR (triangles),  
         The top panel refers to the unpolarized case, while
         the bottom panel to the polarized one.}
\label{theor1}  
\end{figure}

All the effects described in the present section are summarized in 
Fig.\ref{theor1}, where the ratio between the resonance region and the 
'DIS' one is reported for the unpolarized and for the polarized case:
the numerator is obtained from the experimental data, while the denominator 
includes the different components of the present analysis, one by one. 

For unpolarized scattering it has been found that TMC and LxR diminish 
considerably the space left for HT contributions. 
The contribution of TMC is large at the largest values of $Q^2$ because these 
correspond also to large $x$ values. 
Moreover, the effect of TMC is larger than the one of LxR.
The lowest data point at $Q^2\approx 0.4$ GeV$^2$ has been excluded from 
the analysis because of the high uncertainty in both the pQCD calculation and 
the subtraction of TMC. 

Similarly, in polarized scattering the inclusion of TMC and LxR decreases 
the ratio $R_{pol}^{LT}$.
However, in this case these effects are included almost completely 
within the error bars. 
Clearly, duality is strongly violated at $Q^2 < 1.7$ GeV$^2$.

The difference between unpolarized and polarized scattering at low $Q^2$ can 
be attributed {\it e.g.} to unmeasured, so far, $Q^2$ dependent effects, both 
in the asymmetry, $A_1$,  and in $g_2$. Furthermore, a full treatment
of the $Q^2$ dependence would require both a more accurate knowledge of the 
ratio $R$ in the resonance region, and a simultaneous evaluation of $g_2$.
The present mismatch between the unpolarized and polarized low $Q^2$ 
behavior might indicate that factorization is broken differently for the two 
processes, and that the universality of quark descriptions no longer holds. 

\section{Size of Higher Twist (HT) corrections}
\begin{figure}
\resizebox{0.45\textwidth}{!}{%
  \includegraphics{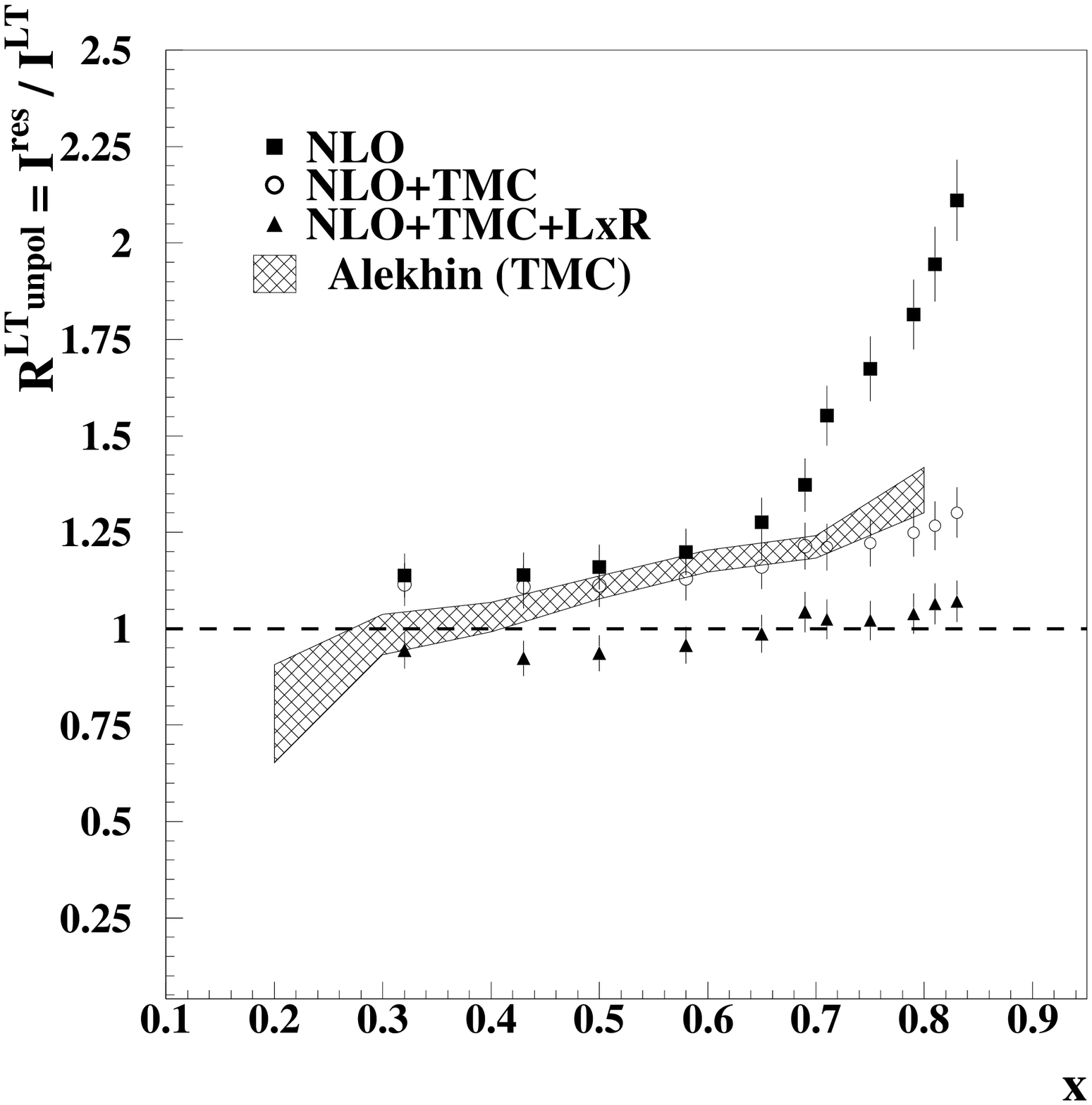}}
\resizebox{0.45\textwidth}{!}{%
  \includegraphics{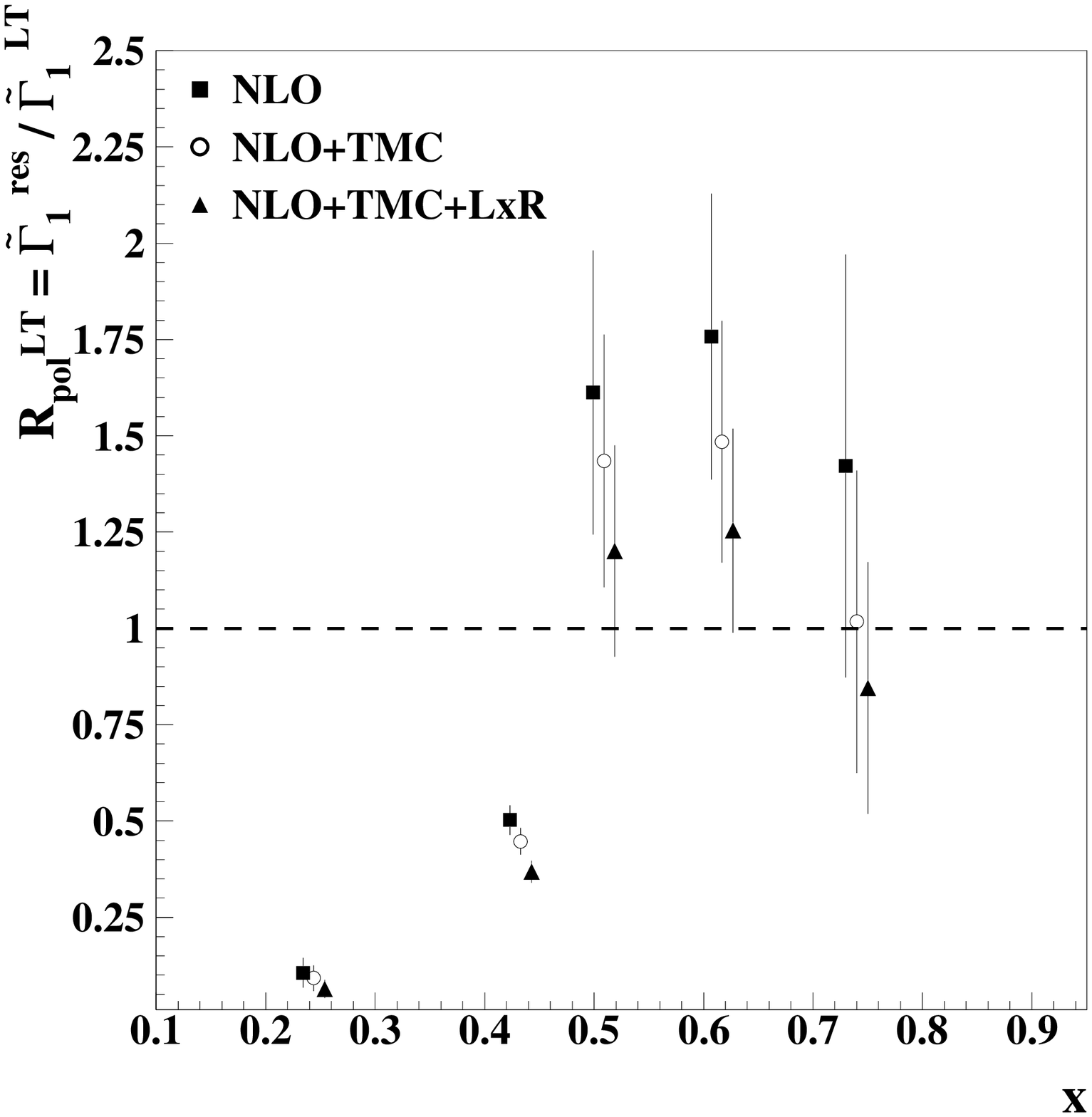}
}
\caption{HT coefficients extracted in the resonance region according 
         to Eq.(\protect\ref{CHT}). Shown in the figure is the quantity
         $1+C_{HT}(x)/Q^2$.
         The top (bottom) panel refers to the unpolarized (polarized) case.}
\label{theorx}      
\end{figure}

\begin{figure}
\resizebox{0.45\textwidth}{!}{%
  \includegraphics{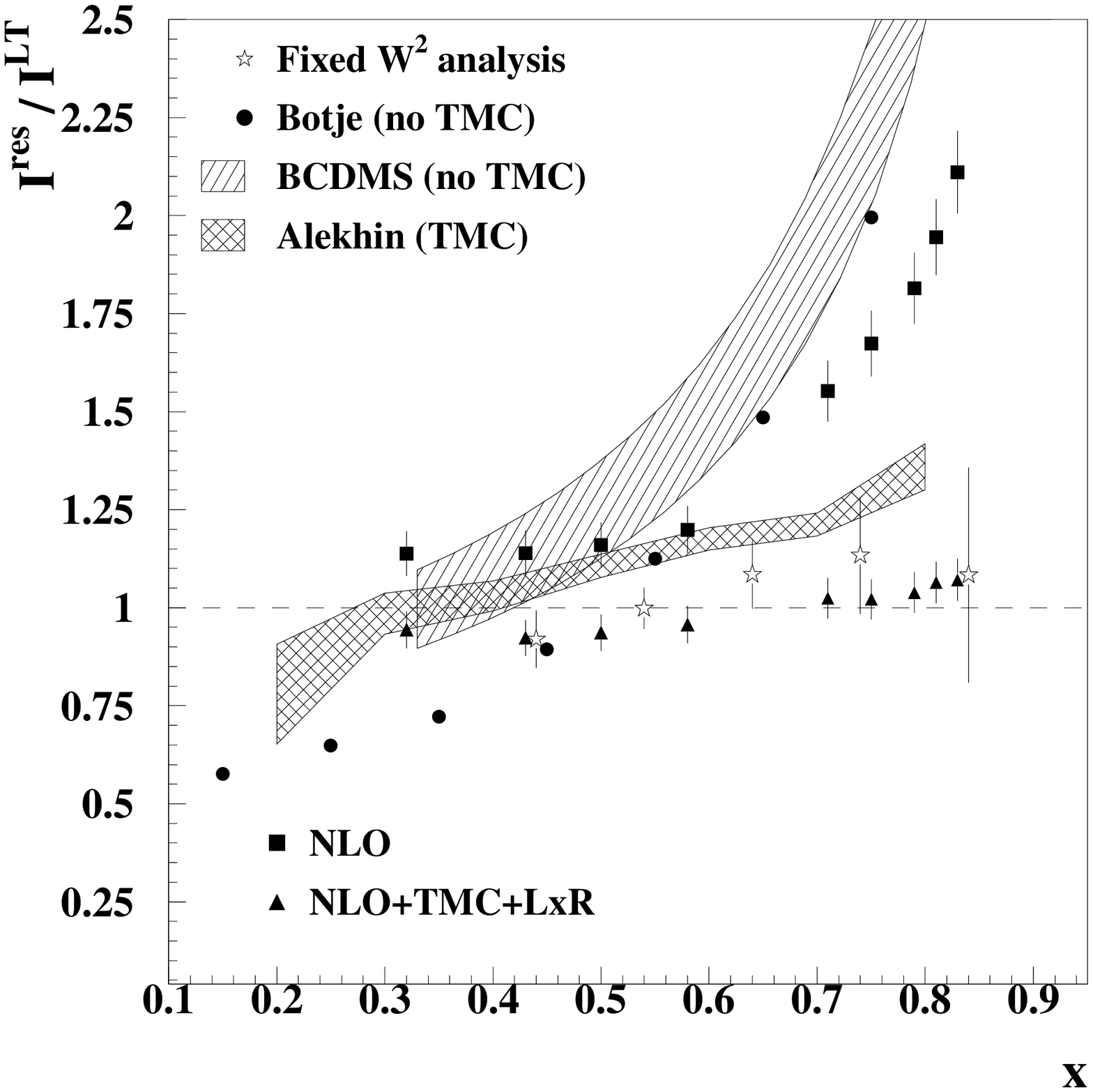}}
\caption{Comparison of the HT coefficient displayed in Fig.3, with other 
         extractions. 
         The triangles and squares are the same as in Fig.3 and they 
         represent the present determination in the resonance region. 
         The results are compared with extractions using DIS data only. 
         The striped hatched area corresponds to the early extraction of 
         Ref.\cite{VM}. The full dots are the central values of the 
         extractions in Refs.\cite{MRST_HT} and \cite{Botje}. 
         These are compared with the more recent extraction of Ref.\cite{Ale1} 
         which includes also TMC. 
         Results obtained in the resonance region, in the fixed $W^2$ analysis 
         of Ref.\cite{SIMO1} are also shown (stars).} 
\label{HTcomp}      
\end{figure}

The discrepancy from unity of the ratios already presented is interpreted in 
terms of HTs.
In Figs.~\ref{theorx},\ref{HTcomp} the question of the size of the HT 
corrections is addressed explicitely. 
For $F_2$, they are defined as:
\begin{equation}
\label{CHT}
H(x,Q^2) = Q^2 \left(F_2^{\mathrm{res}}(x,Q^2) - F_2^{\mathrm{LT}} \right) 
\end{equation}
\begin{equation}
\label{CHT1}
C_{HT}(x) = \frac{H(x,Q^2)}{F_2^{pQCD}(x/Q^2)}
\equiv Q^2 \frac{F_2^{\mathrm{res}}(x,Q^2) - F_2^{LT}}{F_2^{\mathrm{LT}}}
\end{equation}
A similar expression is assumed for $g_1$. 
$C_{HT}$ is the so-called factorized form obtained by assuming that
the $Q^2$ dependences of the LT and of the HT parts are similar and therefore
they cancel out in the ratio. Although the anomalous 
dimensions of the HT part could in principle be different, such a discrepancy 
has not been found so far in accurate analyses of DIS data.   
The HT coefficient, $C_{HT}$ has been evaluated for the three cases listed 
also in Fig.\ref{theor1}, namely with respect to the NLO pQCD calculation, to 
NLO+TMC and to NLO+TMC+LxR. 
The values of $1 + C_{HT}/Q^2$ are plotted in Fig.\ref{theorx} (upper panel) 
as a function of the average value of $x$ 
for each spectrum. One can see that the NLO+TMC+LxR analysis yields 
very small values for $C_{HT}$ in the whole range of $x$. 
Furthermore, the extracted values are consistent 
with the ones obtained in Ref.\cite{SIMO1} using a different method, 
however the present extraction method gives more accurate results. 
Because of the increased precision of our analysis, we are able to disentangle 
the different effects from both TMC and LxR. 
   
In the polarized case (lower panel) the HTs are small within the given 
precision, for $Q^2 > 1.7$ GeV$^2$, but they appear to drop dramatically below 
zero for lower $Q^2$ values. 
The inclusion of TMC and LxR renders these terms consistent with zero at the 
larger $Q^2$ values, but it does not modify substantially their behavior at 
lower $Q^2$. 
It should also be noticed that, by parametrizing the structure functions 
as in Eq.(\ref{CHT}), it is assuming that all of the non-perturbative (np) 
contributions are included in ${\cal O}(1/Q^2)$ twist-4 terms. 
These are in fact the largest type of deviations from a pQCD behavior, 
to be expected at $Q^2$ values of the order of few GeV. 
Only from accurate analyses using a larger number of more precise data,
would one be able to distinguish among different np behaviors.  
From a comparison with results of ratio including phenomenological
parametrizations~\cite{BFL} that includes some of these extra np behaviors 
it's possible to see, however, that their effect seems not be large. 

In Fig.\ref{HTcomp} the results of the present analysis
in the unpolarized case are compared to other current extractions 
of the same quantity. 
These are: {\it i)} the extractions from DIS data, 
performed with the cut $W^2 > 10 $ GeV$^2$ \cite{VM,MRST_HT,Botje}; 
{\it ii)} the recent DIS evaluation by S. Alekhin \cite{Ale1} using 
a cut on $W^2 > 4$ GeV$^2$, and including both TMC and NNLO; {\it iii)} 
the results obtained within a fixed $W^2$ framework~\cite{SIMO1}, 
including both TMC and LxR. 
The results obtained in the deep inelastic region~\cite{ScScSt} also including 
both TMC and LxR yield small HT coefficients, consistent with the ones 
found in Ref.~\cite{SIMO1}.
However, while most of the suppression of the HT in the resonance region is 
attributed to TMC, in \cite{ScScSt} the contribution of TMC is small and the 
suppression is dominated by LxR. 
In other words, the $Q^2$ behavior in the DIS and resonance regions seems to 
be dominated by different effects.   

\section{Conclusions}
A precise and detailed analysis of all publish data in resonance region has
been presented, with the aim of studying the quark-hadron duality in 
unpolarized and polarized $ep$ scattering.
A pQCD NLO analysis including target mass corrections and large $x$ 
resummation effects was extended to the integrals of both unpolarized and 
polarized structure functions in the resonance region. 
Both effects have been quantified and disentangled for the first time.
In the present analysis~\cite{BFL}, duality is satisfied if the pQCD 
calculations agree with the data, modulo higher twist contributions consistent 
with the twist expansion. 
A different behavior for unpolarized and polarized structure functions has
been found, and duality seem strongly violated in the latter case for 
$Q^2<$1.7 GeV$^2$
The discrepancy of the ratio from unity has been interpreted in terms of HTs.
While the size of the HT contributions is comparable in both polarized and 
unpolarized scattering at larger $x$ and $Q^2$ values, at low $x$ and $Q^2$
large negative non-perturbative contributions have been found only in the
polarized case.   
The present detailed extraction of both the $Q^2$ dependence and the HTs in
the resonance region establishes a background for understanding the transition 
between partonic and hadronic degrees of freedom. 
In particular, it seems to be detecting a region where the twist expansion 
breaks down, and at the same time, the data seem to be still far from 
the $Q^2 \rightarrow 0$ limit, where theoretical predictions can be made 
\cite{JiO}.    
More studies addressing this region will be pursued in the future,
some of which are also mentioned in \cite{SIMO1,SIMO_elba}. 
A breakdown of the twist expansion can be interpreted in terms of the dominance
of multi-parton configurations over single parton contributions 
in the scattering process. 
In order to confirm this picture it will be necessary to both extend the
studies of the twist expansion, including the possible $Q^2$ dependence of 
the HT coefficients ~\cite{AF,BFL2} and terms of order ${\cal O}(1/Q^4)$, and
to perform duality studies in semi-inclusive experiments. 

\begin{thebibliography}{}
\bibitem{BG} E.D.~Bloom and F.J.~Gilman, 
             \Journal{\PRL}{25}{1140}{1970};
             \Journal{\PRD}{4}{2901}{1971}.
\bibitem{reticolo} M.~Gockeler {\it et al.}, 
                   \Journal{\PRD}{53}{2317}{1996};
                   LHPC and TXL Coll.,
                   D.~Dolgov {\it et al.},
                   \Journal{\PRD}{66}{034506}{2002}; 
                   W.~Detmold, W.~Melnitchouk \& A.W.~Thomas, 
                   \Journal{\PRD}{66}{054501}{2002}.   
\bibitem{NAC} O.~Nachtmann, \Journal{\NPB}{63}{237}{1973}.
\bibitem{SzcUle} A.~Szczurek \& V.~Uleshchenko, 
                  \Journal{\EPJ}{C12}{663}{2000}. 
\bibitem{BOYA} A.~Bodek \& U.K.~Yang, arXiv:hep-ex/0203009.
\bibitem{CEBAF} I.~Niculescu {\em et al.}, 
                \Journal{\PRL}{85}{1186}{2000}.
\bibitem{whit} L.W.~Whitlow {\em et al.}, \Journal{\PLB}{282}{475}{1992}.
\bibitem{E143} E143 Coll., K.~Abe {\em et al.}, 
               \Journal{\PRD}{58}{112003}{1998}.
\bibitem{HERMESPRL} HERMES Coll., A.~Airapetian {\em et al.}, 
               \Journal{\PRL}{90}{092002}{2003}.
\bibitem{Ale} A.~Fantoni, \Journal{\EPJ}{A17}{385}{2003};
\bibitem{r1990} L.W.~Whitlow {\em et al.}, \Journal{\PLB}{250}{193}{1990}.
\bibitem{E155} E155 Coll., P.L.~Anthony {\em et al.},
               \Journal{\PLB}{493}{19}{2000}. 
\bibitem{BFL} N.~Bianchi, A.~Fantoni \& S.~Liuti,
              \Journal{\PRD}{69}{014505}{2004}.
\bibitem{ALLM} H.~Abramowicz \& A.~Levy, arXiv:hep-ph/9712415.
\bibitem{NMC} NMC Coll., P.~Amaudruz {\em et al.}
                \Journal{\PLB}{364}{107}{1995}.
\bibitem{GeoPol} H.~Georgi \& H.~D.~Politzer,
                 \Journal{\PRD}{14}{1829}{1976}.
\bibitem{Mir} J.~L.~Miramontes \& J.~Sanchez Guillen,
              \Journal{\ZPC}{41}{247}{1988}. 
\bibitem{FraGun} W.~R.~Frazer \& J.~F.~Gunion,
                 \Journal{\PRL}{45}{1138}{1980}.
\bibitem{ioanaPRD} I.~Niculescu, C.~Keppel, S.~Liuti and G.~Niculescu,
Phys.\ Rev.\ D {\bf 60}, 094001 (1999).
\bibitem{SIMO1} S.~Liuti, R.~Ent, C.E.~Keppel \& I.~Niculescu, 
                \Journal{\PRL}{89}{162001}{2002}.
\bibitem{BroLep} S.~J.~Brodsky \& G.~P.~Lepage, SLAC-PUB-2447.
\bibitem{Ama} D.~Amati {\it et al.}, \Journal{\NPB}{173}{429}{1980}.
\bibitem{Rob} R.~G.~Roberts, \Journal{\EPJ}{C10}{697}{1999}.
\bibitem{PenRos} M.~R.~Pennington \& G.~G.~Ross, 
                \Journal{\PLB}{102}{167}{1981}.
\bibitem{VM} M.~Virchaux \& A.~Milsztajn,
             \Journal{\PLB}{274}{221}{1992}.
\bibitem{MRST_HT} A.~D.~Martin {\it et al.}, 
              \Journal{\PLB}{443}{301}{1998}.
\bibitem{Botje} M.~Botje, \Journal{\EPJ}{C14}{285}{2000}.
\bibitem{Ale1} {\bf (a)} S. I. Alekhin, 
             \Journal{\PRD}{63}{094022}{2001};
             \Journal{\PRD}{68}{014002}{2003};
             \Journal{\JHEP}{\bf 02}{015}{2003};
             {\bf (b)} arXiv:hep-ph/0212370.
\bibitem{ScScSt} S.~Schaefer, A.~Schafer \& M.~Stratmann, 
              \Journal{\PLB}{514}{284}{2001}.
\bibitem{JiO} X.~D.~Ji \& J.~Osborne, J.\ Phys.\ G {\bf 27}, 127 (2001).
\bibitem{SIMO_elba} S. Liuti, \Journal{\EPJ}{A17}{385}{2003}. 
\bibitem{AF} A.~Fantoni, Proc. of the workshop ``Structure of the nucleon at 
             large Bjorken $x$'', July 2004, Marseille.
\bibitem{BFL2} N.~Bianchi, A.~Fantoni, S.~Liuti, in preparation.

\end{thebibliography}
\end{document}